\documentclass[sigconf]{acmart}
\DeclareUnicodeCharacter{FF0C}{,}

\AtBeginDocument{%
  \providecommand\BibTeX{{%
    \normalfont B\kern-0.5em{\scshape i\kern-0.25em b}\kern-0.8em\TeX}}}

\usepackage{booktabs}
\usepackage{multirow}
\usepackage{threeparttable}
\usepackage{comment}
\usepackage{subcaption}
\usepackage{xspace}
\usepackage{algorithm}
\usepackage{algorithmic}
\usepackage{amsmath}
\usepackage{hyperref} % 用于 \autoref 命令

\newcommand{\ie}{i.e.\@\xspace}

\newcommand{\etal}{et al.\@\xspace}
\newcommand{\etc}{etc.\@\xspace}

\begin{document}

\title{SimLabel: Similarity-Weighted Iterative Framework for Multi-annotator Learning with Missing Annotations}

\newcommand\approach{SimLabel}

% \affiliation{%

% \author{Anonymous Authors}

% 设置作者信息
\author{Liyun Zhang}
\affiliation{%
  \institution{D3 Center, The University of Osaka}
  \country{Japan}
}
% \email{zhang.liyun@ids.osaka-u.ac.jp}

\author{Zheng Lian}
\affiliation{%
  \institution{Institute of automation, Chinese academy of science}
  \country{China}
}
% \email{lianzheng2016@ia.ac.cn}

\author{Hong Liu}
\affiliation{%
  \institution{Xiamen University}
  \country{China}
}
% \email{lynnliu.xmu@gmail.com}

\author{Takanori Takebe}
\affiliation{%
  \institution{Cincinnati Children's Hospital Medical Center}
  \country{Japan}
}
% \email{takanori.takebe@cchmc.org}

\author{Yuta Nakashima}
\affiliation{%
  \institution{SANKEN, The University of Osaka}
  \country{Japan}
}
% \email{n-yuta@ids.osaka-u.ac.jp}

%% You do not have to enter your paper ID


\begin{abstract}
Multi-annotator learning (MAL) aims to model annotator-specific labeling patterns. However, existing methods face a critical challenge: they simply skip updating annotator-specific model parameters when encountering missing labels—a common scenario in real-world crowdsourced datasets where each annotator labels only small subsets of samples. This leads to inefficient data utilization and overfitting risks. To this end, we propose a novel similarity-weighted semi-supervised learning framework (\approach) that leverages inter-annotator similarities to generate weighted soft labels for missing annotations, enabling the utilization of unannotated samples rather than skipping them entirely. We further introduce a confidence-based iterative refinement mechanism that combines maximum probability with entropy-based uncertainty to prioritize predicted high-quality pseudo-labels to impute missing labels, jointly enhancing similarity estimation and model performance over time. For evaluation, we contribute a new multimodal multi-annotator dataset, AMER2, with high and more variable missing rates, reflecting real-world annotation sparsity and enabling evaluation across different sparsity levels.
% Extensive experiments validate the effectiveness of our method.
\end{abstract}

\begin{CCSXML}
<ccs2012>
   <concept>
       <concept_id>10010147.10010257.10010282.10011305</concept_id>
       <concept_desc>Computing methodologies~Semi-supervised learning settings</concept_desc>
       <concept_significance>500</concept_significance>
       </concept>
   <concept>
       <concept_id>10010147.10010257</concept_id>
       <concept_desc>Computing methodologies~Machine learning</concept_desc>
       <concept_significance>500</concept_significance>
       </concept>
   <concept>
       <concept_id>10010147.10010257.10010258.10010262</concept_id>
       <concept_desc>Computing methodologies~Multi-task learning</concept_desc>
       <concept_significance>500</concept_significance>
       </concept>
 </ccs2012>
\end{CCSXML}

\ccsdesc[500]{Computing methodologies~Semi-supervised learning settings}
\ccsdesc[500]{Computing methodologies~Machine learning}
\ccsdesc[500]{Computing methodologies~Multi-task learning}

% \section{CCS Concepts and User-Defined Keywords}

% Two elements of the ``acmart'' document class provide powerful
% taxonomic tools for you to help readers find your work in an online
% search.

% The ACM Computing Classification System ---
% \url{https://www.acm.org/publications/class-2012} --- is a set of
% classifiers and concepts that describe the computing
% discipline. Authors can select entries from this classification
% system, via \url{https://dl.acm.org/ccs/ccs.cfm}, and generate the
% commands to be included in the \LaTeX\ source.

% User-defined keywords are a comma-separated list of words and phrases
% of the authors' choosing, providing a more flexible way of describing
% the research being presented.

% CCS concepts and user-defined keywords are required for for all
% articles over two pages in length, and are optional for one- and
% two-page articles (or abstracts).

\keywords{Multi-annotator Learning, Missing Labels, Soft Label, Annotator Similarity, Semi-supervision}



% \subsection{The ``Teaser Figure''}
% A ``teaser figure'' is an image, or set of images in one figure, that
% are placed after all author and affiliation information, and before
% the body of the article, spanning the page. If you wish to have such a
% figure in your article, place the command immediately before the
% \verb|\maketitle| command:
% \begin{verbatim}
%   \begin{teaserfigure}
%     \includegraphics[width=\textwidth]{sampleteaser}
%     \caption{figure caption}
%     \Description{figure description}
%   \end{teaserfigure}
% \end{verbatim}

%   \centering
%   \includegraphics[width=\textwidth]{images/overview.jpg}
%   \caption{The application of thermal-to-color image translation }
%   \Description{Introducing our mission with a simulation diagram, so as to show it more vividly.}
%   \label{fig:introduction}

\maketitle

\section{Introduction}
\label{sec:intro}

\begin{figure}[t]
  \centering
   \includegraphics[width=\linewidth]{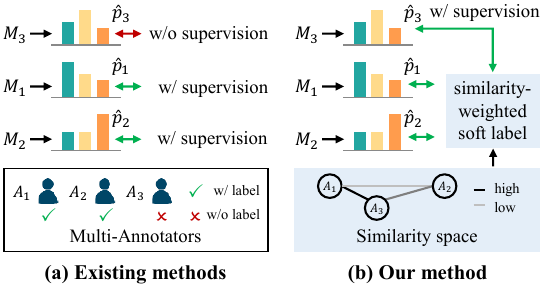}
   \caption{The sample is labeled by annotators $A_1$ to $A_3$, with $A_3$'s label missing. (a) In existing methods, the predicted label distribution $\hat{p}_3$ from $A_3$'s model $M_3$ lacks supervision due to the missing label, resulting in skipped parameter updates for $M_3$ on this sample. (b) In contrast, our method leverages labeling pattern similarities among $A_1$ to $A_3$, estimated from the dataset, to generate a soft label $\bar{p}_3$ that approximates $A_3$'s true label. This is achieved via similarity-weighted aggregation of predictions $\hat{p}_1$ and $\hat{p}_2$, enabling semi-supervised updates of $M_3$ despite label missing.}
   \label{fig:teaser}
\end{figure}

Multi-annotator learning (MAL) has recently emerged as a research hotspot due to its relevance in subjective or nondeterministic tasks, such as medical diagnosis \cite{PADL}, visual perception \cite{CNN-CM}, \etc. MAL aims to model annotator-specific labeling patterns \cite{MaDL, QuMATL}.

However, existing MAL methods face a critical challenge: they simply skip updating annotator-specific model parameters during training when encountering missing labels—a common scenario in real-world crowdsourced datasets, where each annotator labels only a small and often non-overlapping subset of samples to improve annotation efficiency \cite{AMT}. This leads to low data utilization and increased risk of overfitting, as the annotator model is trained on limited annotations due to extensive missing labels.

To address this limitation, we propose a novel similarity-weighted semi-supervised learning framework (\approach), which estimates pairwise inter-annotator similarities and leverages them to generate weighted soft labels for missing annotations. It should be clarified that our goal is not to ``fix'' the inherent missing label characteristic of crowdsourced data, but rather to enable more effective data utilization. We aim to ensure that annotator-specific model parameters can still be updated when labels are missing, rather than simply skipped, thereby improving model performance through enhanced supervision.

Specifically, consider the example in Figure~\ref{fig:teaser} with three annotators ($A_1$, $A_2$, and $A_3$), where $A_3$'s annotation is missing for a given sample. As shown in Figure~\ref{fig:teaser}(a), existing approaches train separate models ($M_1$, $M_2$, and $M_3$) for each annotator. However, when $A_3$'s label is absent, existing methods simply skip updating $M_3$'s parameters entirely due to a lack of supervision, wasting valuable training opportunities and leading to inefficient data utilization and potential overfitting, as repeated application of this practice may result in annotator-specific models being trained on a small dataset with numerous missing labels.

In contrast, our method proposed in Figure~\ref{fig:teaser}(b) leverages Cohen's kappa coefficient \cite{kappa} to calculate pairwise inter-annotator similarities. When $A_3$'s label is missing, we weight the predicted distributions from other annotators ($A_1$ and $A_2$) based on their similarities to $A_3$, generating a similarity-weighted soft label $\bar{p}_3$ to supervise model $M_3$'s training. This semi-supervised approach enables continuous parameter updates rather than skipping missing annotations entirely, thereby improving data utilization efficiency and reducing overfitting risks.

Meanwhile, we introduce a confidence assessment mechanism that combines maximum probability values and entropy-based uncertainty metrics to evaluate the similarity-weighted soft labels. High-confidence predictions exceeding a predetermined threshold represent high reliability of the generated pseudo-labels, which are used to impute original missing labels in the dataset to recalculate the inter-annotator similarity matrix. This establishes a self-reinforcing cycle that jointly enhances similarity estimation and model performance over time.

To facilitate evaluation, we contribute a new multimodal multi-annotator dataset for video emotion recognition, AMER2, with 10 annotators, high and more variable missing rates across annotators (ranging from 75.9\% to 91.3\%). AMER2 better reflects real-world sparse annotation scenarios and enables evaluation under varying levels of label sparsity. Our contributions are as follows:

\begin{itemize}
    \item \textbf{We propose a novel similarity-weighted semi-supervised learning framework} that addresses the missing label challenge in multi-annotator learning. It leverages inter-annotator similarities to generate weighted soft labels, enabling annotator-specific model updates when annotations are missing rather than skipping them entirely. This improves data utilization and reduces overfitting risk, enhancing model performance.

    \item \textbf{We introduce a confidence-based iterative refinement mechanism} that combines maximum probability with entropy-based uncertainty to dynamically prioritize predicted high-quality pseudo-labels to impute missing labels, jointly enhancing similarity estimation and model performance over time.

    \item \textbf{We contribute a new multimodal multi-annotator dataset, AMER2}, with high and more variable missing rates across 10 annotators (ranging from 75.9\% to 91.3\%), which better reflects real-world sparse annotation scenarios and enables evaluation under varying levels of label sparsity.
\end{itemize}
%-------------------------------------------------------------------------

\begin{table*}
\centering
\caption{Label statistics and missing rates of the AMER2 dataset compared to the AMER dataset. For each annotator, the number of labeled samples, the corresponding missing rate (\%), as well as the average data, and the total number of samples are reported. AMER contains 13 annotators while AMER2 contains 10 annotators, denoted as $A_k$.}
\label{tab:dataset_statistic}
\begin{tabular}{lccccccccccccccc}
\toprule
Number of samples & $A_1$ & $A_2$ & $A_3$ & $A_4$ & $A_5$ & $A_6$ & $A_7$ & $A_8$ & $A_9$ & $A_{10}$ & $A_{11}$ & $A_{12}$ & $A_{13}$ & Average & Total \\
\midrule
AMER & 1096 & 1031 & 1022 & 1036 & 1012 & 970 & 1064 & 1049 & 1060 & 1062 & 5187 & 5197 & 5202 & 1999.1 & 5207 \\
AMER2 & 545 & 538 & 201 & 557 & 493 & 545 & 346 & 544 & 542 & 545 & - & - & - & 485.6 & 2311 \\
\midrule
Missing rate (\%) & $A_1$ & $A_2$ & $A_3$ & $A_4$ & $A_5$ & $A_6$ & $A_7$ & $A_8$ & $A_9$ & $A_{10}$ & $A_{11}$ & $A_{12}$ & $A_{13}$ & Average & - \\
\midrule
AMER & 79.0 & 80.2 & 80.4 & 80.1 & 80.6 & 81.4 & 79.6 & 79.9 & 79.6 & 79.6 & 0.4 & 0.2 & 0.1 & 69.6 & - \\
AMER2 & 76.4 & 76.7 & 91.3 & 75.9 & 78.7 & 76.4 & 85.0 & 76.5 & 76.5 & 76.4 & - & - & - & 79.0 & - \\
\bottomrule
\end{tabular}
\end{table*}

\section{Related Work}

\subsection{Traditional Multi-annotator Learning}
Traditional multi-annotator learning aims to estimate a single consensus or ground-truth label from multiple annotators' labels. These include early probabilistic models \cite{DS_model}, EM algorithms \cite{LFC, GLAD}, Gaussian models \cite{GP-MLL}, CNN models \cite{Aggnet}, and biased estimation \cite{bias_annotator}. Tanno et al. \cite{tanno2019learning} proposed modeling annotator confusion matrices as learnable parameters in neural networks. Cao et al. \cite{cao2019max} introduced max-MIG to learn from multiple annotators. NEAL \cite{NEAL} employs neural expectation-maximization to jointly learn annotator expertise and true labels. Later methods used probabilistic frameworks to aggregate multiple annotations into a consensus or ground-truth label by confusion matrix \cite{Sampling-CM}, agreement distribution \cite{Learn2agree}, and Gaussian distributions \cite{PADL}. This aggregation paradigm often treats annotator disagreements as noise to be averaged away rather than valuable information \cite{noise1, Correction}. The underlying computer vision and machine learning techniques used in multi-annotator learning have broader applications across various domains~\cite{Panoptic-tcsvt, Panoptic-wacv, Thermal-to-Color, PhD, 3DFacePolicy, uneven}, though annotator disagreements in multi-annotator scenarios reflect genuine perspective differences rather than noise.

\subsection{Multi-annotator Labeling Pattern Modeling}
Some studies have also attempted to model individual annotator patterns and provide explanations: D-LEMA \cite{D-LEMA} trains annotator models on non-contradictory subsets with spatial weights for noise handling; MaDL \cite{MaDL} jointly optimizes ground truth classifiers and annotator models through weighted embeddings; TAX \cite{TAX} associates convolutional kernels with prototype libraries for pixel-level annotation decisions; MAGI \cite{MAGI} leverages annotator explanations to address noisy annotations, and Schaekermann et al.~\cite{structured_adjudication} analyze factors contributing to disagreements. Particularly, QuMATL \cite{QuMATL} models individual annotator labeling patterns via learnable queries with behavioral analysis, introducing a paradigm shift, \ie, views each annotator as having unique labeling patterns worth preserving rather than as noisy approximations of ground truth. However, missing labels are inherent characteristics of crowdsourced data; the annotator-specific model parameters will skip update in case of missing labels, thereby influencing the individual annotator modeling.

\subsection{Multi-annotator Learning with Missing Labels}
To the best of our knowledge, the multi-annotator learning with missing labels task problem has not yet been investigated. Several similar works are as follows: Yan \etal \cite{yan2014} modeled annotator expertise for complete annotation scenarios without addressing missing annotation challenges. Davani \etal \cite{davani2022} preserved disagreements beyond majority voting but lacked a framework for missing annotator labels. Li \etal \cite{li2021} incorporated instance features but assumed available annotations, while Tanno \etal \cite{tanno2019} addressed noisy labels through confusion estimation without handling missing data. Guan \etal \cite{guan2018} demonstrated individual labeler modeling benefits but did not consider missing label scenarios. Shah \etal \cite{shah2016} explored crowdsourcing self-correction mechanisms without specific missing label strategies. Rodrigues \etal \cite{rodrigues2018} proposed deep learning from crowds assuming complete crowd annotations, and Albarqouni \etal \cite{albarqouni2016} introduced AggNet for medical imaging but focused on aggregating available annotations rather than handling missing ones.

Existing multi-annotator learning methods directly skip annotator-specific model parameter updates in case of missing labels, which not only causes inefficient data utilization but may also lead to overfitting of annotator models on small datasets.
Our work fills this critical gap by proposing a framework that leverages the similarity relationships between annotators to achieve annotator-specific model parameter updates when labels are missing rather than simply skipping. By introducing this similarity-based soft constraint for cases of missing labels, our approach avoids the low data utilization efficiency and potential overfitting risks caused by skipping annotator-specific model parameter updates due to a lack of supervision when labels are missing.

\begin{figure*}[h]
  \centering
  \includegraphics[width=0.76\linewidth]{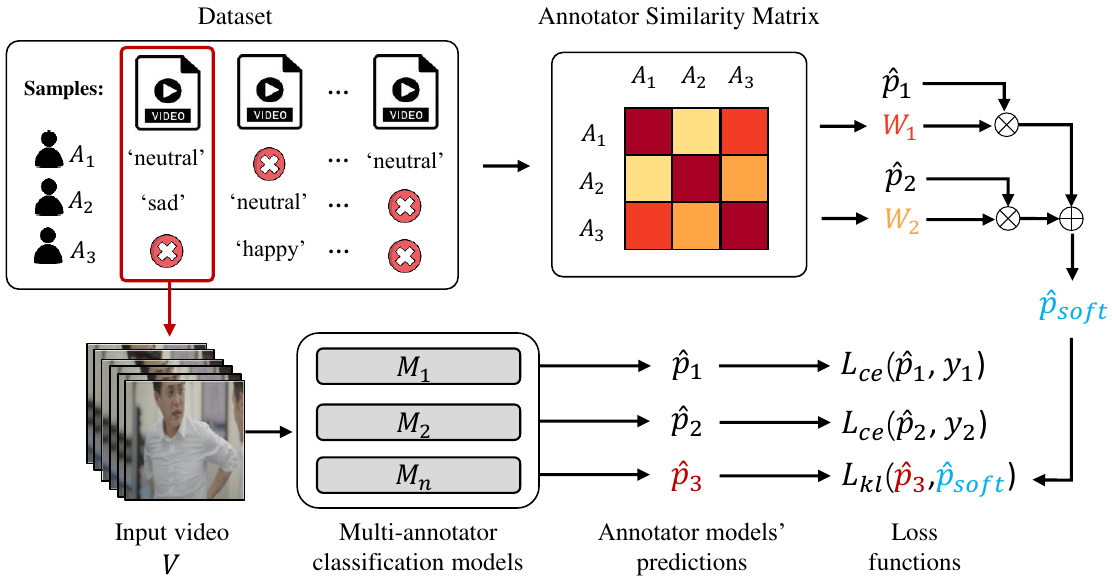}
  \caption{The main framework of \approach. For datasets with missing labels, annotator similarity is calculated via the Cohen's kappa coefficient (darker colors indicate higher similarity). Using video sample $V$ with missing label from $A_n$: Multiple annotator-specific models process $V$ to produce label distributions. Labeled predictions ($\hat{p}_1$, $\hat{p}_2$) are supervised through cross-entropy loss with ground truth labels ($y_1$, $y_2$). Unlabeled predictions ($\hat{p}_n$) is constrained via KL divergence loss with a soft label ($\hat{p}_{soft}$), generated as a similarity-weighted combination of $\hat{p}_1$ and $\hat{p}_2$ using weights $W_1$ and $W_2$ derived from annotator similarities to $A_n$.}
  \label{fig:main}
\end{figure*}

\section{Dataset Construction}
\label{sec:dataset}
This paper introduces a new video emotion recognition dataset, AMER2, which is an extended version of the AMER dataset \cite{QuMATL}. The AMER dataset contains 5,207 video samples and provides rich per-annotator labels to meet the requirements of multi-annotator tendency learning \cite{QuMATL}.

Unlike AMER, AMER2 provides an additional 2,311 samples and sparse per-annotator labels, with the intention of validating the effectiveness of our proposed method under more challenging missing conditions. In AMER2, most samples focus on single-person videos with relatively complete speech content, sourced from movies and TV series. 

During the annotation process, we utilize the Label Studio toolkit \cite{LabelStudio} and hire multiple annotators who are masters or PhD students in our labs. To ensure annotation quality, these annotators first undergo preliminary exams. In these exams, we provide 10 samples and ask the annotators to select the most likely label from 8 candidate labels: \emph{worry}, \emph{happiness}, \emph{neutral}, \emph{anger}, \emph{surprise}, \emph{sadness}, \emph{other}, and \emph{unknown}. These samples were previously annotated by five experts and have obtained five-agreement labels. Annotators who fail to pass the preliminary exam are removed from the annotation pool. After that, we retained 10 annotators, and each annotator completed the task in approximately two weeks, with scheduled breaks to maintain annotation quality. Finally, each annotator provided approximately 201 to 557 labels.

Table \ref{tab:dataset_statistic} provides statistics for AMER and AMER2. From this table, we observe that AMER2 has an overall average missing rate of 79.0\%, with one annotator's missing labels reaching an extreme of 91.30\%, which is higher than AMER's overall average missing rate of 69.6\%. Therefore, AMER2 increases the proportion of missing labels to better mimic real-world scenarios with sparse per-annotator labels. In this paper, we conduct experiments on both datasets, aiming to validate the effectiveness of our method under variable missing rates.  

\section{Methodology}
\label{sec:method}
\begin{figure}[t]
  \centering
  \includegraphics[width=0.9\linewidth]{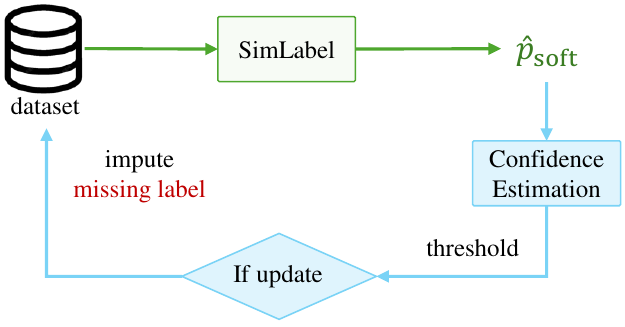}
  \caption{\textbf{Confidence-based Iterative Refinement.} Confidence is calculated for the soft label ($\bar{p}_{k}$) generated for missing labels. When confidence exceeds a predetermined threshold, the predicted high-quality pseudo-label imputes the original missing label in the dataset, iteratively updating the inter-annotator similarity matrix and utilized for soft label ($\bar{p}_{k}$) generation in subsequent iterations.}
  \label{fig:confidence}
\end{figure}

\approach\ contains two components: the similarity-weighted framework and confidence-based iterative refinement. Similarity-weighted framework generates soft labels for missing annotations through inter-annotator similarity weights, providing semi-supervised constraints for annotator-specific models (Figure~\ref{fig:main}). Confidence-based iterative refinement dynamically updates the similarity matrix by evaluating confidence scores of generated soft labels, creating a self-reinforcing learning cycle (Figure~\ref{fig:confidence}).

\subsection{Similarity-weighted Framework}
We propose a novel approach to address the issue of multi-annotator learning with missing labels. As shown in Figure~\ref{fig:main}, our dataset consists of pairs of a video $x$ and a set $\mathcal{Y} = \{y\}$ of labels $y \in \{0, 1\}^N$ in the one-hot representation, where $N$ is the number of classes. $\mathcal{Y}$ contains labels by multiple annotators $A_k$ ($k = 1, \dots, K$, where $K$ is the number of annotators), providing different annotations for the same video because of the subjectivity or nondeterministicity of the task. Often, not all annotators label all samples (thus $|\mathcal{Y}| \leq K)$, resulting in missing labels---a common situtaion in real-world scenarios. In this example, for emotion assessment in the video, $A_1$ gives the label \textit{`neutral'}, $A_2$ gives the label \textit{`sad'}, while $A_3$ does not give the label, i.e, the label is missing.

The video $x$ is processed by separate classification model $M_k$ for each annotator $A_k$, designed to learn individual labeling patterns. The choice of these classification models are arbitrary: They can use Gaussian distribution fitting (PADL \cite{PADL}), confusion matrix (MaDL \cite{MaDL}), or query-based architecture (QuMATL \cite{QuMATL}), etc.~to model individual annotators (and their relationship). Given an input data $x$, each model produces label distribution $\hat{p}_k(l|x)$ for $A_k$ and class $l$ ($l = 1,\dots,N$).

When the annotation from $A_k$ is available for pair $(x, \mathcal{Y}$, we update the corresponding model $M_k$ using $y_k \in \mathcal{Y}$ through supervised learning by computing cross-entropy loss:
\begin{equation}
\mathcal{L}_{\text{ce}}(\hat{p}_k, y_k) = -\sum_{l = 1}^N y_{kl}\log \hat{p}_k(l|x).
\end{equation}

When annotation from $A_k$ is unavailable, we update model $M_k$ in a semi-supervised manner by computing Kullback-Leibler (KL) divergence loss with generating a soft label $\bar{p}_k \in [0, 1]^N$:
\begin{equation}
\mathcal{L}_{\text{kl}}(\hat{p}_k, \bar{p}_k) = D_{\mathrm{KL}}(\hat{p}_k \| \bar{p}_k).
\end{equation}

The soft label $\bar{p}_k$ is key to our method. Based on our assumption that inter-annotator correlations derived from labels in a multi-annotator dataset can give some ideas on missing labels, inter-annotator similarities across the entire dataset to indirectly update models with unannotated samples through semi-supervised learning.
We calculate the similarity matrix between annotators using the Cohen's kappa coefficient \cite{kappa} from the original dataset. Figure~\ref{fig:main} illustrates the matrix, where darker colors indicate greater similarity between annotators. $A_1$ has higher similarity to $A_3$ compared to $A_2$; therefore, when $A_3$'s label is missing, $A_1$'s label $y_1$ may be more informative to predict $y_3$ compared $A_2$'s. We define the similarity weights of $A_1$ relative to $A_3$ and $A_2$ relative to $A_3$ as $w_{1,3}$ and $w_{2,3}$, respectively. The soft label $\bar{p}_{3}$ is thus generated by the weighted sum of label distribution predictions $\hat{p}_1$ and $\hat{p}_2$ with their corresponding similarity weights $w_{1,3}$ and $w_{2,3}$. In general, we generate the soft label for $A_k$ by:
\begin{equation}
\bar{p}_{k} = \sum w_{k', k} \hat{p}_{k'} ,
\end{equation}
where the summation is computed over all $k' \not= k$. Here, $k'$ refers to the number of those annotators who have labels; $k$ refers to the annotator who has the missing label. 

% The total training loss $\mathcal{L}_{\text{total}}$ for our multi-annotator classification model is defined as:
% \begin{equation}
% \mathcal{L}_{\text{total}} = \sum_{k=1}^{K} \mathcal{L}_{\text{ce}}(\hat{p}_k, y_k) + \mathcal{L}_{\text{kl}}(\hat{p}_k, \bar{p}_{k}).
% \end{equation}
%-------------------------------------------------------------------------

\begin{algorithm}[t]
\caption{The Confidence-based Iterative Refinement for Dynamic Inter-Annotator Similarity Relationship}
\label{alg:similarity}
\begin{algorithmic}[1]
\REQUIRE Dataset $\mathcal{D}$ with missing labels, confidence threshold $T$
\STATE Initialize similarity matrix $SM$ using Cohen's kappa coefficient on available labels
\STATE Initialize annotator models $\{M_1, M_2, \ldots, M_K\}$
\FOR{each training epoch}
    \FOR{each sample with missing labels}
        \STATE Generate similarity-weighted soft label: \\
        \hspace{1cm}$\bar{p}_{k} = \sum w_{k', k} \hat{p}_{k'}$
        \STATE Calculate confidence: \\
        \hspace{1cm}$c = \max(\bar{p}_{k}) \times (1 - \frac{H[\bar{p}_{k}]}{H_\text{max}})$
        \IF{$c \geq T$}
            \STATE Extract predicted label: \\
            \hspace{1cm}$y_{pred} = \arg\max(\bar{p}_{k})$
            \STATE Impute $y_{pred}$ into dataset for corresponding missing annotation
            \STATE Recalculate annotator similarity matrix $SM$ using updated dataset
        \ENDIF
        \STATE Update annotator models using supervised and semi-supervised losses
    \ENDFOR
\ENDFOR
\ENSURE Refined similarity matrix $SM$, imputed dataset, trained annotator models
\end{algorithmic}
\end{algorithm}

\subsection{Confidence-based Iterative Refinement}
Building upon the similarity-weighted framework, we introduce a confidence assessment mechanism for the similarity-weighted soft labels. As shown in Figure~\ref {fig:confidence}, if the confidence score exceeds a predetermined threshold, indicating high reliability of the generated soft label, the predicted label will impute the corresponding missing labels of the original dataset to recalculate the inter-annotator similarity matrix. Through this mechanism, we establish a self-reinforcing cycle that continuously refines the inter-annotator similarity relationships and the individual annotator models' ability, leading to progressively more accurate predictions throughout the training process. This mechanism integrates both maximum probability values and entropy-based uncertainty metrics to provide comprehensive confidence estimates and identify highly reliable soft labels.

{\bf Confidence Calculation.}
As shown in Algorithm~\ref{alg:similarity}, for the similarity-weighted soft labels $\bar{p}_{k}$ generated for a missing annotation of $A_k$, we perform confidence calculations to enhance our semi-supervised method. Our confidence combines maximum probability with normalized entropy to provide a comprehensive assessment of prediction reliability:
\begin{equation}
c = \max(\bar{p}_{k}) \times (1 - \frac{H[\bar{p}_{k}]}{H_\text{max}}),
\end{equation}
where $\max(\bar{p}_{k})$ is the maximum value in $\bar{p}_{k}$, $H[\bar{p}_{k}]$ is the entropy of $\bar{p}_{k}$, and $H_\text{max} = \log N$. The right side of the multiplication is to normalize $c$ into $[0, 1]$. 

This formulation requires predictions to have both high maximum probability and low normalized entropy to achieve high confidence scores. This provides a comprehensive assessment of prediction reliability by balancing two key factors: (1) The maximum probability term $\max(\bar{p}_{k})$ captures the model's confidence in the most likely class. (2) The normalized entropy term $H_{norm}(\bar{p}_{k}) = (1 - \frac{H[\bar{p}_{k}]}{H_\text{max}})$ measures the uncertainty across the entire distribution, with lower values indicating more concentrated (certain) predictions.

{\bf Dynamic Refinement Process.}
As shown in Algorithm~\ref{alg:similarity}, when the calculated confidence exceeds a predetermined threshold $T$ (Algorithm~\ref{alg:similarity}, line 7), indicating that the generated soft label has high reliability, the predicted label $y_{pred}$ is extracted and incorporated into the location of missing label in the original dataset (lines 8--9). The annotator similarity matrix $SM$ is then recalculated using the updated dataset with newly imputed labels (line 10).

This process facilitates more accurate establishment of similarity relationships between annotators in cases of missing labels. As training progresses and missing labels meeting confidence criteria are incorporated, a virtuous cycle emerges, continuously refining the similarity relationships between annotators. The dynamic refinement allows each annotator model to more accurately capture its specific annotation patterns, even when starting from datasets with significant numbers of missing annotations.

The effectiveness of this approach lies in its ability to leverage high-confidence predictions to bootstrap the learning process, creating a self-improving system where each iteration potentially enhances the quality of both the similarity matrix and the generated soft labels for remaining missing annotations.
% Specifically, for the similarity-weighted soft labels $\hat{p}_{soft}$ generated for missing annotations, we perform additional confidence calculations to enhance this semi-supervised approach. This confidence measure combines maximum probability with minimum entropy, where entropy considers the entire probability distribution to better reflect prediction uncertainty, resulting in a more comprehensive confidence assessment. If the confidence exceeds a predetermined threshold $T$, indicating that the generated soft label has high reliability, it is incorporated into the original dataset to recalculate the annotator similarity matrix $SM$.
% This process aims to facilitate more accurate establishment of similarity relationships between annotators in cases of missing labels. As training progresses and missing labels meeting confidence criteria are incorporated, a virtuous cycle emerges, continuously refining the similarity relationships between annotators and helping individual annotator models more accurately capture their specific annotation patterns.
%-------------------------------------------------------------------------

\begin{table*}[t]
\centering
% \small
\caption{Accuracy comparison on AMER2 dataset (10 annotators, $A_{k}$, \textit{k} = 1, \dots, 10) for annotator modeling performance with average (Avg). Methods compared: existing approach (Architecture - Skip); similarity-weighted framework (Architecture - Ours); similarity-weighted framework with confidence-based iterative refinement (Architecture - Ours + Confidence).}
\label{tab:accuracy1}
\begin{tabular}{llccccccccccc}
\toprule
Methods & $A_1$ & $A_2$ & $A_3$ & $A_4$ & $A_5$ & $A_6$ & $A_7$ & $A_8$ & $A_9$ & $A_{10}$ & Avg \\
\midrule
PADL - Skip & 0.81 & 0.84 & 0.82 & 0.87 & 0.78 & 0.80 & 0.78 & 0.84 & 0.80 & 0.79 & 0.81  \\
PADL - Ours & 0.84 & 0.85 & \textbf{0.84} & 0.89 & 0.82 & 0.81 & 0.82 & 0.85 & 0.82 & 0.83 & 0.84  \\
PADL - Ours + Confidence & \textbf{0.86} & \textbf{0.87} & \textbf{0.84} & \textbf{0.90} & \textbf{0.84} & \textbf{0.83} & \textbf{0.85} & \textbf{0.86} & \textbf{0.85} & \textbf{0.86} & \textbf{0.86}  \\
% \cmidrule{1-1}
\midrule
MaDL - Skip & 0.84 & 0.83 & 0.82 & 0.86 & 0.81 & 0.82 & 0.80 & 0.82 & 0.85 & 0.84 & 0.83  \\
MaDL - Ours & 0.87 & 0.84 & 0.85 & 0.87 & 0.83 & 0.85 & 0.82 & 0.86 & 0.86 & 0.85 & 0.85  \\
MaDL - Ours + Confidence & \textbf{0.89} & \textbf{0.86} & \textbf{0.88} & \textbf{0.88} & \textbf{0.86} & \textbf{0.87} & \textbf{0.84} & \textbf{0.88} & \textbf{0.88} & \textbf{0.87} & \textbf{0.87}  \\
% \cmidrule{1-1}
\midrule
QuMATL - Skip & 0.86 & 0.83 & 0.87 & 0.84 & 0.86 & 0.87 & 0.88 & 0.83 & 0.85 & 0.86 & 0.86 \\
QuMATL - Ours & \textbf{0.89} & 0.85 & 0.90 & 0.86 & 0.89 & 0.88 & 0.91 & \textbf{0.86} & 0.87 & 0.89 & 0.88 \\
QuMATL - Ours + Confidence & \textbf{0.89} & \textbf{0.87} & \textbf{0.92} & \textbf{0.88} & \textbf{0.91} & \textbf{0.90} & \textbf{0.93} & \textbf{0.86} & \textbf{0.90} & \textbf{0.91} & \textbf{0.90} \\
\bottomrule
\end{tabular}
\end{table*}

\begin{table*}[t]
\centering
% \small
\caption{Accuracy comparison on AMER dataset evaluating annotator modeling performance for 13 annotators.}
% ($A_{k}$, \textit{k} = 1, \dots, 13) and average performance (Avg). Methods compared: existing approach (Architecture - Skip), our similarity-weighted framework (Architecture - Ours), and similarity-weighted framework with confidence-based iterative refinement (Architecture - Ours + Confidence).}
\label{tab:accuracy2}
\begin{tabular}{l@{\hspace{2mm}}lccccccccccccc}
\toprule
Methods & $A_1$ & $A_2$ & $A_3$ & $A_4$ & $A_5$ & $A_6$ & $A_7$ & $A_8$ & $A_9$ & $A_{10}$ & $A_{11}$ & $A_{12}$ & $A_{13}$ & Avg \\
\midrule
PADL - Skip & 0.89 & 0.90 & 0.88 & 0.93 & 0.87 & 0.91 & 0.86 & 0.94 & 0.89 & 0.88 & 0.47 & 0.54 & 0.35 & 0.79  \\
PADL - Ours & 0.91 & 0.92 & 0.90 & \textbf{0.94} & 0.90 & 0.92 & 0.89 & 0.95 & 0.91 & 0.91 & 0.55 & 0.61 & 0.45 & 0.83  \\
PADL - Ours + Confidence & \textbf{0.92} & \textbf{0.93} & \textbf{0.91} & \textbf{0.94} & \textbf{0.91} & \textbf{0.93} & \textbf{0.90} & \textbf{0.96} & \textbf{0.92} & \textbf{0.93} & \textbf{0.59} & \textbf{0.65} & \textbf{0.50} & \textbf{0.85}  \\
% \cmidrule{1-1}
\midrule
MaDL - Skip & 0.93 & 0.91 & 0.90 & 0.89 & 0.90 & 0.88 & 0.90 & 0.89 & 0.87 & 0.92 & 0.50 & 0.53 & 0.37 & 0.80  \\
MaDL - Ours & 0.95 & 0.92 & 0.92 & 0.91 & 0.92 & 0.90 & 0.92 & 0.92 & 0.90 & \textbf{0.94} & 0.59 & 0.60 & 0.48 & 0.84  \\
MaDL - Ours + Confidence & \textbf{0.96} & \textbf{0.93} & \textbf{0.93} & \textbf{0.92} & \textbf{0.93} & \textbf{0.91} & \textbf{0.93} & \textbf{0.93} & \textbf{0.91} & \textbf{0.94} & \textbf{0.64} & \textbf{0.65} & \textbf{0.53} & \textbf{0.86}  \\
% \cmidrule{1-1}
\midrule
QuMATL - Skip & 0.94 & 0.93 & 0.93 & 0.94 & 0.94 & 0.92 & 0.93 & 0.95 & 0.93 & 0.93 & 0.59 & 0.61 & 0.40 & 0.84 \\
QuMATL - Ours & 0.96 & 0.94 & \textbf{0.95} & 0.95 & 0.95 & 0.94 & 0.95 & 0.96 & 0.95 & 0.95 & 0.68 & 0.69 & 0.52 & 0.88 \\
QuMATL - Ours + Confidence & \textbf{0.97} & \textbf{0.95} & \textbf{0.95} & \textbf{0.96} & \textbf{0.96} & \textbf{0.95} & \textbf{0.96} & \textbf{0.97} & \textbf{0.96} & \textbf{0.96} & \textbf{0.72} & \textbf{0.73} & \textbf{0.57} & \textbf{0.89} \\
\bottomrule
\end{tabular}
\end{table*}

\begin{table*}[t]
\centering
% \small
\caption{Randomly removing annotations at 40\% missing ratios is to simulate sparser missing scenarios on STREET, AMER, and AMER2 datasets.  -Ha, -He, -Sa, -Li, and -Or represent five perspectives: happiness, healthiness, safety, liveliness, and orderliness. The average modeling performance of whole annotators is evaluated by the accuracy metric.}
% \caption{Randomly removing annotations at different missing ratios (20\%, 30\%, and 40\%) is to simulate label missing scenarios on the STREET dataset,  -Ha, -He, -Sa, -Li, and -Or represent five perspectives of STREET dataset: happiness, healthiness, safety, liveliness, and orderliness. Here, we only show the average performance of 10 annotators' modeling. This is also applied to AMER2 and AMER datasets to increase the missing rates to validate our approach. Here, the accuracy metric is used for evaluation. Higher is better.}
\label{tab:accuracy3}
\begin{tabular}{l@{\hspace{2.1mm}}ccccccc}
\toprule
Methods & STREET-Ha & STREET-He & STREET-Sa & STREET-Li & STREET-Or & AMER & AMER2 \\
\midrule
PADL - Skip & 0.43 & 0.42 & 0.38 & 0.41 & 0.40 & 0.58 & 0.61 \\
PADL - Ours & 0.48 & 0.46 & 0.42 & 0.47 & 0.45 & \textbf{0.64} & 0.66 \\
PADL - Ours + Confidence & \textbf{0.51} & \textbf{0.49} & \textbf{0.45} & \textbf{0.50} & \textbf{0.48} & \textbf{0.64} & \textbf{0.69} \\
% \cmidrule{1-1}
\midrule
MaDL - Skip & 0.42 & 0.43 & 0.36 & 0.38 & 0.36 & 0.63 & 0.65 \\
MaDL - Ours & \textbf{0.45} & 0.46 & 0.38 & 0.41 & 0.39 & 0.66 & 0.69 \\
MaDL - Ours + Confidence & \textbf{0.45} & \textbf{0.48} & \textbf{0.41} & \textbf{0.43} & \textbf{0.42} & \textbf{0.71} & \textbf{0.72} \\
% \cmidrule{1-1}
\midrule
QuMATL - Skip & 0.52 & 0.51 & 0.46 & 0.49 & 0.48 & 0.66 & 0.68 \\
QuMATL - Ours & 0.56 & 0.55 & 0.50 & 0.53 & 0.52 & 0.70 & 0.71 \\
QuMATL - Ours + Confidence & \textbf{0.60} & \textbf{0.58} & \textbf{0.54} & \textbf{0.57} & \textbf{0.56} & \textbf{0.74} & \textbf{0.75} \\
% Full - PADL - Skip & 0.58 & 0.59 & 0.52 & 0.58 & 0.55 & 0.87 & 0.87 \\
% Full - PADL - Ours & 0.58 & 0.59 & 0.52 & 0.58 & 0.55 & 0.87 & 0.87 \\
% Full - MaDL - Skip & 0.58 & 0.59 & 0.52 & 0.58 & 0.55 & 0.87 & 0.87 \\
% Full - MaDL - Ours & 0.58 & 0.59 & 0.52 & 0.58 & 0.55 & 0.87 & 0.87 \\
% Full - QuMATL - Skip & 0.63 & 0.64 & 0.58 & 0.62 & 0.61 & 0.92 & 0.92 \\
% Full - QuMATL - Ours & 0.63 & 0.64 & 0.58 & 0.62 & 0.61 & 0.92 & 0.92 \\
\bottomrule
\end{tabular}
\end{table*}

\begin{table}[!t]
\centering
% \small
\caption{DIC measures how well enhanced annotator modeling via missing label handling improves inter-annotator consistency toward ground truth, lower values indicate better gains. -S and -O represent Skip and Ours approaches.}
\label{tab:consistency}
\begin{tabular}{c@{\hspace{2.5mm}}c@{\hspace{2.5mm}}c@{\hspace{2.5mm}}c@{\hspace{2.5mm}}c}
\toprule
Datasets & PADL-S & PADL-O & QuMATL-S & QuMATL-O \\
\midrule
STREET-Ha & 0.48 & \textbf{0.44} & 0.43 & \textbf{0.38} \\
STREET-He & 0.52 & \textbf{0.45} & 0.38 & \textbf{0.34} \\
STREET-Sa & 0.32 & \textbf{0.28} & 0.24 & \textbf{0.20} \\
STREET-Li & 0.43 & \textbf{0.38} & 0.27 & \textbf{0.22} \\
STREET-Or & 0.57 & \textbf{0.53} & 0.54 & \textbf{0.49} \\
AMER & 0.36 & \textbf{0.32} & 0.23 & \textbf{0.19} \\
AMER2 & 0.34 & \textbf{0.31} & 0.22 & \textbf{0.17} \\
\bottomrule
\end{tabular}
\end{table}

\section{Experiment}
\label{sec:experiment}
We conduct extensive experiments comparing \approach\ (with and without confidence-based iterative refinement) against existing approaches that simply skip annotator-specific model parameter updates in case of missing labels. To verify effectiveness across different architectures, we evaluate on three annotator modeling frameworks: Gaussian distribution fitting (PADL \cite{PADL}), confusion matrix (MaDL \cite{MaDL}), and query-based modeling (QuMATL \cite{QuMATL}). Experiments are conducted on AMER2 and AMER containing real missing labels, and the STREET dataset with simulated missing labels at various missing ratios, using Accuracy and Difference of Inter-annotator Consistency (DIC) \cite{QuMATL} as evaluation metrics.

Note that due to space limitations, we discuss additional key issues with experimental results, including training dynamics, threshold sensitivity to missing rates, and strategies for handling noisy labels and avoiding propagation errors, \etc, in the \textbf{supplementary material}.
% We conduct extensive experiments to compare our \approach\ with existing approaches, i.e., directly skipping training on annotator models in case of missing labels, focusing on individual annotator modeling and inter-annotator consistency. To verify the effectiveness of \approach, we select different annotator model architectures: Gaussian distribution fitting architecture (PADL \cite{PADL}), confusion matrix architecture (MaDL \cite{MaDL}), and query-based architecture (QuMATL \cite{QuMATL}). We evaluate and discuss results on the new dataset AMER2, other datasets (AMER and STREET) using accuracy and Difference of Inter-annotator Consistency (DIC) metrics \cite{QuMATL}.

\subsection{Implementation Details}
We use Cohen's kappa coefficient \cite{kappa} to calculate the inter-annotator similarity matrix. Image and video data are all resized to 224$\times$224 and further normalized. For different annotator model architectures, we follow their original training and testing settings. These experiments are achieved on four NVIDIA V100 GPUs.
% For datasets in the experiment, the input image and video are resized to 224$\times$224 and further normalized. The different annotator model architectures all follow their original settings. During training, we use the AdamW optimizer with an initial learning rate of 1e-4, weight decay of 0.01, and a maximum gradient norm of 1.0 for gradient clipping. A linear warmup strategy is applied for the first 20\% steps followed by cosine learning rate decay. We set the maximum number of epochs to 200, with early stopping (patience being 25) to prevent overfitting. To accelerate training, the model is trained using distributed data parallel (DDP) on four NVIDIA V100 GPUs.

\subsection{Evaluation Metrics}
Accuracy is a standard metric to evaluate individual annotator modeling. DIC \cite{QuMATL} quantifies how inter-annotator correlations differ between ground-truths and predictions, and we also use DIC to evaluate our approach's benefits from the perspective of inter-annotator consistency.

\subsection{Datasets}
For the dataset, we primarily utilize the newly constructed multimodal emotion recognition dataset AMER2, alongside the earlier version AMER and the city impression assessment dataset STREET \cite{QuMATL}. AMER2 and AMER naturally contain missing labels in real-world settings, while STREET is a complete real-world dataset. The AMER dataset's complexity in capturing temporal emotion dynamics aligns with recent advances in time-sensitive emotion recognition~\cite{MicroEmo-arxiv, MicroEmo-mm}. For the STREET dataset, we randomly remove annotations at different missing ratios to simulate annotation absence. Similarly, we also apply this random removal procedure to AMER2 and AMER datasets to further increase missing rates and validate the effectiveness of our approach.

\subsection{Results Analysis}
Table~\ref{tab:accuracy1} and Table~\ref{tab:accuracy2} present accuracy results for each annotator across different annotator model architectures based on the comparison between our proposed \approach\ (\ie, using only the similarity-weighted framework of the similarity-weighted soft label, defined as ``- Ours'', and using both the similarity-weighted framework and confidence-based iterative refinement, defined as ``- Ours + Confidence'') with existing approaches (\ie, directly skipping annotator-specific model parameter updates in case of missing labels, defined as ``- Skip''). Table~\ref{tab:accuracy3} presents average accuracy results for multi-annotators across different annotator model architectures at different missing labels.

On the AMER2 dataset (Table~\ref{tab:accuracy1}), our approach using the similarity-weighted framework (``- Ours'') consistently outperforms existing approaches (``- Skip'') which directly skip annotator-specific model parameter updates in case of missing labels, with average improvements of 3\% for PADL, 2\% for MaDL, and 2\% for QuMATL. When incorporating confidence-based iterative refinement (``- Ours + Confidence'') for our approach, we observe further enhancements of 2\% across all different architectures. 

Results on the AMER dataset (Table~\ref{tab:accuracy2}) show similar patterns of improvement. Our approach using the similarity-weighted framework (``- Ours'') improves average accuracy by around 3\% for PADL, MaDL, and QuMATL compared to existing approaches (``- Skip''). With confidence-based iterative refinement (``- Ours + Confidence'') for our approach, these improvements further increase to around 2\%. This consistent enhancement across different model architectures of multi-annotator learning validates our central hypothesis that leveraging inter-annotator similarity provides an effective framework for addressing missing label challenges in multi-annotator learning.

For the STREET dataset, we conducted experiments with artificially induced missing labels at different rates, we show 40\% (Table~\ref{tab:accuracy3}) here, and more data are provided in the supplementary material. To evaluate robustness, we also apply this random removal procedure to AMER2 and AMER datasets to further increase missing rates and validate the effectiveness of our approach. Our approach delivers consistent improvements across all scenarios, which demonstrates that our method is particularly valuable in scenarios with severe label sparsity.
% with average gains of around 4-7\% depending on the dataset and missing ratio. 

% Importantly, the relative improvement becomes more pronounced as the missing ratio increases, with average gains of 7\% at 40\% missing labels compared to 5-6\% at 20\%. This demonstrates that our method is particularly valuable in scenarios with severe label sparsity. On the STREET dataset with its five assessment perspectives, our approach shows uniform improvements, with the largest gains observed for the more challenging safety perspective (Sa).

For the gain evaluation of inter-annotator consistency, the DIC scores in Table~\ref{tab:consistency} also show that our similarity-weighted approach shows consistent gains compared to the skip way (\ie, skipping annotator-specific model parameter updates in case of missing labels) across different architectures on different datasets.
Lower DIC values indicate that our approach better captures individual annotators' labeling patterns through enhanced annotator modeling via missing label handling, thereby improving inter-annotator consistency convergence toward ground truth. More detailed Table data is provided in the supplementary material.

These results consistently demonstrate that leveraging annotator similarity relationships through our soft label generation and confidence-based iterative refinement mechanism improves multi-annotator modeling performance, especially in realistic scenarios with missing annotations.
%-------------------------------------------------------------------------

\begin{table}[t]
\centering
% \small
\caption{Ablation studies on similarity matrix calculation, confidence threshold, and calculation choices on AMER2 dataset. Top: Similarity matrix calculation choice. Middle: Performance with different confidence thresholds. Bottom: Comparison of different confidence calculation methods.}
\label{tab:ablation}
\begin{tabular}{cccc}
\toprule
Similarity matrix calculation & PADL & MaDL & QuMATL \\
\midrule
Pearson correlation & 0.82 & 0.84 & 0.86 \\
Krippendorff's alpha & 0.84 & 0.85 & 0.88 \\
Cohen's kappa & \textbf{0.85} & \textbf{0.87} & \textbf{0.90} \\
\midrule
Confidence Threshold & PADL & MaDL & QuMATL \\
\midrule
$\tau=0.5$ & 0.85 & 0.86 & 0.89 \\
$\tau=0.6$ & \textbf{0.86} & \textbf{0.87} & \textbf{0.90} \\
$\tau=0.7$ & 0.84 & 0.85 & 0.88 \\
$\tau=0.8$ & 0.82 & 0.83 & 0.86 \\
% \midrule
% Missing Rate & PADL & MaDL & QuMATL & Best \\
% \midrule
% 20\% & 0.65 & 0.6 & 0.6 & 0.6-0.65 \\
% 30\% & 0.6 & 0.6 & 0.65 & 0.6-0.65 \\
% 40\% & 0.55 & 0.55 & 0.6 & 0.55-0.6 \\
\midrule
Confidence Calculation & PADL & MaDL & QuMATL \\
\midrule
$\max(\bar{p}_{k})$ & 0.83 & 0.84 & 0.87 \\
$(1 - \frac{H[\bar{p}_{k}]}{H_\text{max}})$ & 0.84 & 0.85 & 0.88 \\
$\max(\bar{p}_{k}) \times (1 - \frac{H[\bar{p}_{k}]}{H_\text{max}})$ & \textbf{0.86} & \textbf{0.87} & \textbf{0.90} \\
\bottomrule
\end{tabular}
\end{table}

\subsection{Ablation Study}
To evaluate the design choices in our confidence-based iterative refinement mechanism, we conduct a detailed ablation study examining confidence threshold selection, its sensitivity to different missing rates, and the effectiveness of different confidence formulation methods. They are all performed on AMER2 dataset.

\textbf{Similarity Matrix Calculation.}
We first need to clarify a key point: our confidence threshold does not ``filter out erroneous pseudo-labels'', but rather uses reliable (high-confidence) predictions to progressively refine the similarity matrix. Therefore, under the widely validated Cohen's kappa coefficient \cite{kappa} and self-iterative framework, the model performance demonstrates robustness. Second, we conducted ablation experiments comparing Cohen's kappa with Pearson correlation coefficient \cite{PCC} and Krippendorff's alpha coefficient \cite{KAC} to evaluate the accuracy of the similarity matrix, as shown in Table~\ref{tab:ablation} (top).
Results on the AMER2 dataset show Cohen's kappa consistently outperforms alternatives through chance agreement correction for categorical annotations. Pearson correlation coefficient fails to capture discrete characteristics, while Krippendorff's alpha shows instability in sparse scenarios. Even with these less-matched metrics, model performance degradation is minimal, validating similarity matrix robustness.

\textbf{Confidence Threshold Selection.}
Table~\ref{tab:ablation} (middle) shows the performance of our method with different confidence thresholds. The results indicate that a threshold of $\tau=0.6$ achieves the best performance across all model architectures. Higher thresholds ($\tau=0.8$) lead to performance degradation, likely because too few predictions meet the criteria for updating the similarity matrix. Lower thresholds ($\tau=0.5$) also perform slightly worse than $\tau=0.6$, possibly due to the inclusion of lower-quality predictions that introduce noise into the update process.

\textbf{Confidence Formulation Comparison.}
Finally, we evaluate different confidence calculation methods (Table~\ref{tab:ablation}, bottom): (1) using only maximum probability $\max(\bar{p}_{k})$, (2) using only normalized entropy complement $(1 - \frac{H[\bar{p}_{k}]}{H_\text{max}})$, and (3) our proposed combined approach $\max(\bar{p}_{k}) \times (1 - \frac{H[\bar{p}_{k}]}{H_\text{max}})$, where $\bar{p}_{k}$ represents the similarity-weighted soft label probability distribution generated for missing annotations. The results demonstrate that our combined method consistently outperforms single-metric approaches across all architectures, with an average performance improvement of 3\% over $\max(\bar{p}_{k})$ and 2\% over entropy-only formulation. This validates our hypothesis that effective confidence assessment should consider both the strength of the dominant class prediction and the overall distribution shape.
%-------------------------------------------------------------------------

\section{Conclusion}
We addressed the challenge of missing labels in multi-annotator learning through a similarity-weighted semi-supervised framework that leverages inter-annotator relationships instead of skipping annotator-specific model parameter updates in case of missing labels. \approach\ combines soft label generation with a confidence-based iterative refinement mechanism to dynamically refine inter-annotator similarity estimates. We also contribute a new dataset, AMER2, with high and variable missing rates to reflect real-world annotation sparsity and enable evaluation across different sparsity levels. Extensive experiments on different datasets and missing rates validated \approach's effectiveness to address the missing label challenge. In future work, we plan to extend this framework to handle dynamic annotator behaviors and explore more complex scenarios.
% subjective annotation domains such as medical imaging.
%-------------------------------------------------------------------------

\bibliographystyle{ACM-Reference-Format}
\bibliography{sample-base}


\begin{thebibliography}{41}

%%% the Web address.

\ifx \showCODEN    \undefined \def \showCODEN     #1{\unskip}     \fi
\ifx \showDOI      \undefined \def \showDOI       #1{#1}\fi
\ifx \showISBNx    \undefined \def \showISBNx     #1{\unskip}     \fi
\ifx \showISBNxiii \undefined \def \showISBNxiii  #1{\unskip}     \fi
\ifx \showISSN     \undefined \def \showISSN      #1{\unskip}     \fi
\ifx \showLCCN     \undefined \def \showLCCN      #1{\unskip}     \fi
\ifx \shownote     \undefined \def \shownote      #1{#1}          \fi
\ifx \showarticletitle \undefined \def \showarticletitle #1{#1}   \fi
\ifx \showURL      \undefined \def \showURL       {\relax}        \fi
% invisible to TeX
\providecommand\bibfield[2]{#2}
\providecommand\bibinfo[2]{#2}
\providecommand\natexlab[1]{#1}
\providecommand\showeprint[2][]{arXiv:#2}

\bibitem[Albarqouni et~al\mbox{.}(2016a)]%
        {Aggnet}
\bibfield{author}{\bibinfo{person}{Shadi Albarqouni}, \bibinfo{person}{Christoph Baur}, \bibinfo{person}{Felix Achilles}, \bibinfo{person}{Vasileios Belagiannis}, \bibinfo{person}{Stefanie Demirci}, {and} \bibinfo{person}{Nassir Navab}.} \bibinfo{year}{2016}\natexlab{a}.
\newblock \showarticletitle{Aggnet: deep learning from crowds for mitosis detection in breast cancer histology images}.
\newblock \bibinfo{journal}{\emph{IEEE transactions on medical imaging}} \bibinfo{volume}{35}, \bibinfo{number}{5} (\bibinfo{year}{2016}), \bibinfo{pages}{1313--1321}.
\newblock


\bibitem[Albarqouni et~al\mbox{.}(2016b)]%
        {albarqouni2016}
\bibfield{author}{\bibinfo{person}{Shadi Albarqouni}, \bibinfo{person}{Christoph Baur}, \bibinfo{person}{Felix Achilles}, \bibinfo{person}{Vasileios Belagiannis}, \bibinfo{person}{Stefanie Demirci}, {and} \bibinfo{person}{Nassir Navab}.} \bibinfo{year}{2016}\natexlab{b}.
\newblock \showarticletitle{Aggnet: deep learning from crowds for mitosis detection in breast cancer histology images}.
\newblock \bibinfo{journal}{\emph{IEEE transactions on medical imaging}} \bibinfo{volume}{35}, \bibinfo{number}{5} (\bibinfo{year}{2016}), \bibinfo{pages}{1313--1321}.
\newblock


\bibitem[Benesty et~al\mbox{.}(2009)]%
        {PCC}
\bibfield{author}{\bibinfo{person}{Jacob Benesty}, \bibinfo{person}{Jingdong Chen}, \bibinfo{person}{Yiteng Huang}, {and} \bibinfo{person}{Israel Cohen}.} \bibinfo{year}{2009}\natexlab{}.
\newblock \showarticletitle{Pearson correlation coefficient}.
\newblock In \bibinfo{booktitle}{\emph{Noise reduction in speech processing}}. \bibinfo{publisher}{Springer}, \bibinfo{pages}{1--4}.
\newblock


\bibitem[Cao et~al\mbox{.}(2019)]%
        {cao2019max}
\bibfield{author}{\bibinfo{person}{Peng Cao}, \bibinfo{person}{Yilun Xu}, \bibinfo{person}{Yuqing Kong}, {and} \bibinfo{person}{Yizhou Wang}.} \bibinfo{year}{2019}\natexlab{}.
\newblock \showarticletitle{Max-mig: an information theoretic approach for joint learning from crowds}.
\newblock \bibinfo{journal}{\emph{arXiv preprint arXiv:1905.13436}} (\bibinfo{year}{2019}).
\newblock


\bibitem[Chen et~al\mbox{.}(2023)]%
        {NEAL}
\bibfield{author}{\bibinfo{person}{Junfan Chen}, \bibinfo{person}{Richong Zhang}, \bibinfo{person}{Jie Xu}, \bibinfo{person}{Chunming Hu}, {and} \bibinfo{person}{Yongyi Mao}.} \bibinfo{year}{2023}\natexlab{}.
\newblock \showarticletitle{A Neural Expectation-Maximization Framework for Noisy Multi-Label Text Classification}.
\newblock \bibinfo{journal}{\emph{IEEE Transactions on Knowledge and Data Engineering}} \bibinfo{volume}{35}, \bibinfo{number}{11} (\bibinfo{year}{2023}), \bibinfo{pages}{10992--11003}.
\newblock
\urldef\tempurl%
\url{https://doi.org/10.1109/TKDE.2022.3223067}
\showDOI{\tempurl}


\bibitem[Cheng et~al\mbox{.}(2023)]%
        {TAX}
\bibfield{author}{\bibinfo{person}{Yuan-Chia Cheng}, \bibinfo{person}{Zu-Yun Shiau}, \bibinfo{person}{Fu-En Yang}, {and} \bibinfo{person}{Yu-Chiang~Frank Wang}.} \bibinfo{year}{2023}\natexlab{}.
\newblock \showarticletitle{TAX: Tendency-and-Assignment Explainer for Semantic Segmentation with Multi-Annotators}.
\newblock \bibinfo{journal}{\emph{arXiv preprint arXiv:2302.09561}} (\bibinfo{year}{2023}).
\newblock


\bibitem[Davani et~al\mbox{.}(2022)]%
        {davani2022}
\bibfield{author}{\bibinfo{person}{Aida~Mostafazadeh Davani}, \bibinfo{person}{Mark D{\'\i}az}, {and} \bibinfo{person}{Vinodkumar Prabhakaran}.} \bibinfo{year}{2022}\natexlab{}.
\newblock \showarticletitle{Dealing with disagreements: Looking beyond the majority vote in subjective annotations}.
\newblock \bibinfo{journal}{\emph{Transactions of the Association for Computational Linguistics}}  \bibinfo{volume}{10} (\bibinfo{year}{2022}), \bibinfo{pages}{92--110}.
\newblock


\bibitem[Dawid and Skene(1979)]%
        {DS_model}
\bibfield{author}{\bibinfo{person}{Alexander~Philip Dawid} {and} \bibinfo{person}{Allan~M Skene}.} \bibinfo{year}{1979}\natexlab{}.
\newblock \showarticletitle{Maximum likelihood estimation of observer error-rates using the EM algorithm}.
\newblock \bibinfo{journal}{\emph{Journal of the Royal Statistical Society: Series C (Applied Statistics)}} \bibinfo{volume}{28}, \bibinfo{number}{1} (\bibinfo{year}{1979}), \bibinfo{pages}{20--28}.
\newblock


\bibitem[Guan et~al\mbox{.}(2018)]%
        {guan2018}
\bibfield{author}{\bibinfo{person}{Melody Guan}, \bibinfo{person}{Varun Gulshan}, \bibinfo{person}{Andrew Dai}, {and} \bibinfo{person}{Geoffrey Hinton}.} \bibinfo{year}{2018}\natexlab{}.
\newblock \showarticletitle{Who said what: Modeling individual labelers improves classification}. In \bibinfo{booktitle}{\emph{Proceedings of the AAAI conference on artificial intelligence}}, Vol.~\bibinfo{volume}{32}.
\newblock


\bibitem[Gwet(2011)]%
        {KAC}
\bibfield{author}{\bibinfo{person}{Kilem~L Gwet}.} \bibinfo{year}{2011}\natexlab{}.
\newblock \showarticletitle{On the Krippendorff’s alpha coefficient}.
\newblock \bibinfo{journal}{\emph{Manuscript submitted for publication. Retrieved October}} \bibinfo{volume}{2}, \bibinfo{number}{2011} (\bibinfo{year}{2011}), \bibinfo{pages}{2011}.
\newblock


\bibitem[Herde et~al\mbox{.}(2023)]%
        {MaDL}
\bibfield{author}{\bibinfo{person}{Marek Herde}, \bibinfo{person}{Denis Huseljic}, {and} \bibinfo{person}{Bernhard Sick}.} \bibinfo{year}{2023}\natexlab{}.
\newblock \showarticletitle{Multi-annotator Deep Learning: A Probabilistic Framework for Classification}.
\newblock \bibinfo{journal}{\emph{arXiv preprint arXiv:2304.02539}} (\bibinfo{year}{2023}).
\newblock


\bibitem[Jinadu et~al\mbox{.}(2023)]%
        {Correction}
\bibfield{author}{\bibinfo{person}{Uthman Jinadu}, \bibinfo{person}{Jesse Annan}, \bibinfo{person}{Shanshan Wen}, {and} \bibinfo{person}{Yi Ding}.} \bibinfo{year}{2023}\natexlab{}.
\newblock \showarticletitle{Loss Modeling for Multi-Annotator Datasets}.
\newblock \bibinfo{journal}{\emph{arXiv preprint arXiv:2311.00619}} (\bibinfo{year}{2023}).
\newblock


\bibitem[Li et~al\mbox{.}(2021)]%
        {li2021}
\bibfield{author}{\bibinfo{person}{Jingzheng Li}, \bibinfo{person}{Hailong Sun}, \bibinfo{person}{Jiyi Li}, \bibinfo{person}{Zhijun Chen}, \bibinfo{person}{Renshuai Tao}, {and} \bibinfo{person}{Yufei Ge}.} \bibinfo{year}{2021}\natexlab{}.
\newblock \showarticletitle{Learning from multiple annotators by incorporating instance features}.
\newblock \bibinfo{journal}{\emph{arXiv preprint arXiv:2106.15146}} (\bibinfo{year}{2021}).
\newblock


\bibitem[Liao et~al\mbox{.}(2024)]%
        {PADL}
\bibfield{author}{\bibinfo{person}{Zehui Liao}, \bibinfo{person}{Shishuai Hu}, \bibinfo{person}{Yutong Xie}, {and} \bibinfo{person}{Yong Xia}.} \bibinfo{year}{2024}\natexlab{}.
\newblock \showarticletitle{Modeling annotator preference and stochastic annotation error for medical image segmentation}.
\newblock \bibinfo{journal}{\emph{Medical Image Analysis}}  \bibinfo{volume}{92} (\bibinfo{year}{2024}), \bibinfo{pages}{103028}.
\newblock


\bibitem[Mirikharaji et~al\mbox{.}(2021)]%
        {D-LEMA}
\bibfield{author}{\bibinfo{person}{Zahra Mirikharaji}, \bibinfo{person}{Kumar Abhishek}, \bibinfo{person}{Saeed Izadi}, {and} \bibinfo{person}{Ghassan Hamarneh}.} \bibinfo{year}{2021}\natexlab{}.
\newblock \showarticletitle{D-lema: Deep learning ensembles from multiple annotations-application to skin lesion segmentation}. In \bibinfo{booktitle}{\emph{Proceedings of the IEEE/CVF Conference on Computer Vision and Pattern Recognition}}. \bibinfo{pages}{1837--1846}.
\newblock


\bibitem[Paolacci et~al\mbox{.}(2010)]%
        {AMT}
\bibfield{author}{\bibinfo{person}{Gabriele Paolacci}, \bibinfo{person}{Jesse Chandler}, {and} \bibinfo{person}{Panagiotis~G. Ipeirotis}.} \bibinfo{year}{2010}\natexlab{}.
\newblock \showarticletitle{Running experiments on Amazon Mechanical Turk}.
\newblock \bibinfo{journal}{\emph{Judgment and Decision Making}} \bibinfo{volume}{5}, \bibinfo{number}{5} (\bibinfo{year}{2010}), \bibinfo{pages}{411–419}.
\newblock
\urldef\tempurl%
\url{https://doi.org/10.1017/S1930297500002205}
\showDOI{\tempurl}


\bibitem[Raykar et~al\mbox{.}(2010)]%
        {LFC}
\bibfield{author}{\bibinfo{person}{Vikas~C. Raykar}, \bibinfo{person}{Shipeng Yu}, \bibinfo{person}{Linda~H. Zhao}, \bibinfo{person}{Gerardo~Hermosillo Valadez}, \bibinfo{person}{Charles Florin}, \bibinfo{person}{Luca Bogoni}, {and} \bibinfo{person}{Linda Moy}.} \bibinfo{year}{2010}\natexlab{}.
\newblock \showarticletitle{Learning From Crowds}.
\newblock \bibinfo{journal}{\emph{Journal of Machine Learning Research}} \bibinfo{volume}{11}, \bibinfo{number}{43} (\bibinfo{year}{2010}), \bibinfo{pages}{1297--1322}.
\newblock
\urldef\tempurl%
\url{http://jmlr.org/papers/v11/raykar10a.html}
\showURL{%
\tempurl}


\bibitem[Rodrigues and Pereira(2018)]%
        {rodrigues2018}
\bibfield{author}{\bibinfo{person}{Filipe Rodrigues} {and} \bibinfo{person}{Francisco Pereira}.} \bibinfo{year}{2018}\natexlab{}.
\newblock \showarticletitle{Deep learning from crowds}. In \bibinfo{booktitle}{\emph{Proceedings of the AAAI conference on artificial intelligence}}, Vol.~\bibinfo{volume}{32}.
\newblock


\bibitem[Rodrigues et~al\mbox{.}(2014)]%
        {GP-MLL}
\bibfield{author}{\bibinfo{person}{Filipe Rodrigues}, \bibinfo{person}{Francisco Pereira}, {and} \bibinfo{person}{Bernardete Ribeiro}.} \bibinfo{year}{2014}\natexlab{}.
\newblock \showarticletitle{Gaussian Process Classification and Active Learning with Multiple Annotators}. In \bibinfo{booktitle}{\emph{Proceedings of the 31st International Conference on Machine Learning}} \emph{(\bibinfo{series}{Proceedings of Machine Learning Research}, Vol.~\bibinfo{volume}{32})}, \bibfield{editor}{\bibinfo{person}{Eric~P. Xing} {and} \bibinfo{person}{Tony Jebara}} (Eds.). \bibinfo{publisher}{PMLR}, \bibinfo{address}{Bejing, China}, \bibinfo{pages}{433--441}.
\newblock
\urldef\tempurl%
\url{https://proceedings.mlr.press/v32/rodrigues14.html}
\showURL{%
\tempurl}


\bibitem[Schaekermann et~al\mbox{.}(2019)]%
        {structured_adjudication}
\bibfield{author}{\bibinfo{person}{Mike Schaekermann}, \bibinfo{person}{Graeme Beaton}, \bibinfo{person}{Minahz Habib}, \bibinfo{person}{Andrew Lim}, \bibinfo{person}{Kate Larson}, {and} \bibinfo{person}{Edith Law}.} \bibinfo{year}{2019}\natexlab{}.
\newblock \showarticletitle{Understanding Expert Disagreement in Medical Data Analysis through Structured Adjudication}.
\newblock \bibinfo{journal}{\emph{Proc. ACM Hum.-Comput. Interact.}} \bibinfo{volume}{3}, \bibinfo{number}{CSCW}, Article \bibinfo{articleno}{76} (\bibinfo{date}{Nov.} \bibinfo{year}{2019}), \bibinfo{numpages}{23}~pages.
\newblock
\urldef\tempurl%
\url{https://doi.org/10.1145/3359178}
\showDOI{\tempurl}


\bibitem[Sha et~al\mbox{.}(2024)]%
        {3DFacePolicy}
\bibfield{author}{\bibinfo{person}{Xuanmeng Sha}, \bibinfo{person}{Liyun Zhang}, \bibinfo{person}{Tomohiro Mashita}, {and} \bibinfo{person}{Yuki Uranishi}.} \bibinfo{year}{2024}\natexlab{}.
\newblock \showarticletitle{3DFacePolicy: Speech-Driven 3D Facial Animation with Diffusion Policy}.
\newblock \bibinfo{journal}{\emph{arXiv preprint arXiv:2409.10848}} (\bibinfo{year}{2024}).
\newblock


\bibitem[Shah and Zhou(2016)]%
        {shah2016}
\bibfield{author}{\bibinfo{person}{Nihar Shah} {and} \bibinfo{person}{Dengyong Zhou}.} \bibinfo{year}{2016}\natexlab{}.
\newblock \showarticletitle{No oops, you won’t do it again: mechanisms for self-correction in crowdsourcing}. In \bibinfo{booktitle}{\emph{International conference on machine learning}}. PMLR, \bibinfo{pages}{1--10}.
\newblock


\bibitem[Tanno et~al\mbox{.}(2019a)]%
        {tanno2019learning}
\bibfield{author}{\bibinfo{person}{Ryutaro Tanno}, \bibinfo{person}{Ardavan Saeedi}, \bibinfo{person}{Swami Sankaranarayanan}, \bibinfo{person}{Daniel~C. Alexander}, {and} \bibinfo{person}{Nathan Silberman}.} \bibinfo{year}{2019}\natexlab{a}.
\newblock \showarticletitle{Learning From Noisy Labels by Regularized Estimation of Annotator Confusion}. In \bibinfo{booktitle}{\emph{Proceedings of the IEEE/CVF Conference on Computer Vision and Pattern Recognition (CVPR)}}.
\newblock


\bibitem[Tanno et~al\mbox{.}(2019b)]%
        {Sampling-CM}
\bibfield{author}{\bibinfo{person}{Ryutaro Tanno}, \bibinfo{person}{Ardavan Saeedi}, \bibinfo{person}{Swami Sankaranarayanan}, \bibinfo{person}{Daniel~C Alexander}, {and} \bibinfo{person}{Nathan Silberman}.} \bibinfo{year}{2019}\natexlab{b}.
\newblock \showarticletitle{Learning from noisy labels by regularized estimation of annotator confusion}. In \bibinfo{booktitle}{\emph{Proceedings of the IEEE/CVF conference on computer vision and pattern recognition}}. \bibinfo{pages}{11244--11253}.
\newblock


\bibitem[Tanno et~al\mbox{.}(2019c)]%
        {tanno2019}
\bibfield{author}{\bibinfo{person}{Ryutaro Tanno}, \bibinfo{person}{Ardavan Saeedi}, \bibinfo{person}{Swami Sankaranarayanan}, \bibinfo{person}{Daniel~C Alexander}, {and} \bibinfo{person}{Nathan Silberman}.} \bibinfo{year}{2019}\natexlab{c}.
\newblock \showarticletitle{Learning from noisy labels by regularized estimation of annotator confusion}. In \bibinfo{booktitle}{\emph{Proceedings of the IEEE/CVF conference on computer vision and pattern recognition}}. \bibinfo{pages}{11244--11253}.
\newblock


\bibitem[Tkachenko et~al\mbox{.}(2025)]%
        {LabelStudio}
\bibfield{author}{\bibinfo{person}{Maxim Tkachenko}, \bibinfo{person}{Mikhail Malyuk}, \bibinfo{person}{Andrey Holmanyuk}, {and} \bibinfo{person}{Nikolai Liubimov}.} \bibinfo{year}{2020-2025}\natexlab{}.
\newblock \bibinfo{title}{{Label Studio}: Data labeling software}.
\newblock
\newblock
\urldef\tempurl%
\url{https://github.com/HumanSignal/label-studio}
\showURL{%
\tempurl}
\newblock
\shownote{Open source software available from https://github.com/HumanSignal/label-studio}.


\bibitem[Viera and Garrett(2005)]%
        {kappa}
\bibfield{author}{\bibinfo{person}{AnthonyJ. Viera} {and} \bibinfo{person}{JoanneM. Garrett}.} \bibinfo{year}{2005}\natexlab{}.
\newblock \showarticletitle{Understanding interobserver agreement: the kappa statistic.}
\newblock \bibinfo{journal}{\emph{Family Medicine,Family Medicine}} (\bibinfo{date}{May} \bibinfo{year}{2005}).
\newblock


\bibitem[Wang et~al\mbox{.}(2023)]%
        {Learn2agree}
\bibfield{author}{\bibinfo{person}{Chongyang Wang}, \bibinfo{person}{Yuan Gao}, \bibinfo{person}{Chenyou Fan}, \bibinfo{person}{Junjie Hu}, \bibinfo{person}{Tin~Lum Lam}, \bibinfo{person}{Nicholas~D Lane}, {and} \bibinfo{person}{Nadia Bianchi-Berthouze}.} \bibinfo{year}{2023}\natexlab{}.
\newblock \showarticletitle{Learn2agree: Fitting with multiple annotators without objective ground truth}. In \bibinfo{booktitle}{\emph{International Workshop on Trustworthy Machine Learning for Healthcare}}. Springer, \bibinfo{pages}{147--162}.
\newblock


\bibitem[Welinder et~al\mbox{.}(2010)]%
        {bias_annotator}
\bibfield{author}{\bibinfo{person}{Peter Welinder}, \bibinfo{person}{Steve Branson}, \bibinfo{person}{Pietro Perona}, {and} \bibinfo{person}{Serge Belongie}.} \bibinfo{year}{2010}\natexlab{}.
\newblock \showarticletitle{The multidimensional wisdom of crowds}.
\newblock \bibinfo{journal}{\emph{Advances in neural information processing systems}}  \bibinfo{volume}{23} (\bibinfo{year}{2010}).
\newblock


\bibitem[Whitehill et~al\mbox{.}(2009)]%
        {GLAD}
\bibfield{author}{\bibinfo{person}{Jacob Whitehill}, \bibinfo{person}{Ting-fan Wu}, \bibinfo{person}{Jacob Bergsma}, \bibinfo{person}{Javier Movellan}, {and} \bibinfo{person}{Paul Ruvolo}.} \bibinfo{year}{2009}\natexlab{}.
\newblock \showarticletitle{Whose vote should count more: Optimal integration of labels from labelers of unknown expertise}.
\newblock \bibinfo{journal}{\emph{Advances in neural information processing systems}}  \bibinfo{volume}{22} (\bibinfo{year}{2009}).
\newblock


\bibitem[Yan et~al\mbox{.}(2014a)]%
        {noise1}
\bibfield{author}{\bibinfo{person}{Yan Yan}, \bibinfo{person}{R{\'o}mer Rosales}, \bibinfo{person}{Glenn Fung}, \bibinfo{person}{Ramanathan Subramanian}, {and} \bibinfo{person}{Jennifer Dy}.} \bibinfo{year}{2014}\natexlab{a}.
\newblock \showarticletitle{Learning from multiple annotators with varying expertise}.
\newblock \bibinfo{journal}{\emph{Machine learning}}  \bibinfo{volume}{95} (\bibinfo{year}{2014}), \bibinfo{pages}{291--327}.
\newblock


\bibitem[Yan et~al\mbox{.}(2014b)]%
        {yan2014}
\bibfield{author}{\bibinfo{person}{Yan Yan}, \bibinfo{person}{R{\'o}mer Rosales}, \bibinfo{person}{Glenn Fung}, \bibinfo{person}{Ramanathan Subramanian}, {and} \bibinfo{person}{Jennifer Dy}.} \bibinfo{year}{2014}\natexlab{b}.
\newblock \showarticletitle{Learning from multiple annotators with varying expertise}.
\newblock \bibinfo{journal}{\emph{Machine learning}}  \bibinfo{volume}{95} (\bibinfo{year}{2014}), \bibinfo{pages}{291--327}.
\newblock


\bibitem[Zhang(2024)]%
        {MicroEmo-arxiv}
\bibfield{author}{\bibinfo{person}{Liyun Zhang}.} \bibinfo{year}{2024}\natexlab{}.
\newblock \showarticletitle{MicroEmo: Time-Sensitive Multimodal Emotion Recognition with Micro-Expression Dynamics in Video Dialogues}.
\newblock \bibinfo{journal}{\emph{arXiv preprint arXiv:2407.16552}} (\bibinfo{year}{2024}).
\newblock


\bibitem[Zhang et~al\mbox{.}(2025)]%
        {QuMATL}
\bibfield{author}{\bibinfo{person}{Liyun Zhang}, \bibinfo{person}{Zheng Lian}, \bibinfo{person}{Hong Liu}, \bibinfo{person}{Takanori Takebe}, {and} \bibinfo{person}{Yuta Nakashima}.} \bibinfo{year}{2025}\natexlab{}.
\newblock \showarticletitle{QuMATL: Query-based Multi-annotator Tendency Learning}.
\newblock \bibinfo{journal}{\emph{arXiv preprint arXiv:2503.15237}} (\bibinfo{year}{2025}).
\newblock


\bibitem[Zhang et~al\mbox{.}(2014)]%
        {uneven}
\bibfield{author}{\bibinfo{person}{Liyun Zhang}, \bibinfo{person}{Nanyan Liu}, \bibinfo{person}{Yuanbin Hou}, {and} \bibinfo{person}{Xiaojian Liu}.} \bibinfo{year}{2014}\natexlab{}.
\newblock \showarticletitle{Uneven illumination image segmentation based on multi-threshold S-F}.
\newblock \bibinfo{journal}{\emph{Opto-Electronic Engineering}} \bibinfo{volume}{41}, \bibinfo{number}{7} (\bibinfo{year}{2014}), \bibinfo{pages}{81--87}.
\newblock


\bibitem[Zhang et~al\mbox{.}(2024)]%
        {MicroEmo-mm}
\bibfield{author}{\bibinfo{person}{Liyun Zhang}, \bibinfo{person}{Zhaojie Luo}, \bibinfo{person}{Shuqiong Wu}, {and} \bibinfo{person}{Yuta Nakashima}.} \bibinfo{year}{2024}\natexlab{}.
\newblock \showarticletitle{MicroEmo: Time-Sensitive Multimodal Emotion Recognition with Subtle Clue Dynamics in Video Dialogues}. In \bibinfo{booktitle}{\emph{Proceedings of the 2nd International Workshop on Multimodal and Responsible Affective Computing}}. \bibinfo{pages}{110--115}.
\newblock


\bibitem[Zhang et~al\mbox{.}(2023b)]%
        {Panoptic-tcsvt}
\bibfield{author}{\bibinfo{person}{Liyun Zhang}, \bibinfo{person}{Photchara Ratsamee}, \bibinfo{person}{Zhaojie Luo}, \bibinfo{person}{Yuki Uranishi}, \bibinfo{person}{Manabu Higashida}, {and} \bibinfo{person}{Haruo Takemura}.} \bibinfo{year}{2023}\natexlab{b}.
\newblock \showarticletitle{Panoptic-level image-to-image translation for object recognition and visual odometry enhancement}.
\newblock \bibinfo{journal}{\emph{IEEE Transactions on Circuits and Systems for Video Technology}} \bibinfo{volume}{34}, \bibinfo{number}{2} (\bibinfo{year}{2023}), \bibinfo{pages}{938--954}.
\newblock


\bibitem[Zhang et~al\mbox{.}(2022)]%
        {Thermal-to-Color}
\bibfield{author}{\bibinfo{person}{Liyun Zhang}, \bibinfo{person}{Photchara Ratsamee}, \bibinfo{person}{Yuki Uranishi}, \bibinfo{person}{Manabu Higashida}, {and} \bibinfo{person}{Haruo Takemura}.} \bibinfo{year}{2022}\natexlab{}.
\newblock \showarticletitle{Thermal-to-Color Image Translation for Enhancing Visual Odometry of Thermal Vision}. In \bibinfo{booktitle}{\emph{2022 IEEE International Symposium on Safety, Security, and Rescue Robotics (SSRR)}}. IEEE, \bibinfo{pages}{33--40}.
\newblock


\bibitem[Zhang et~al\mbox{.}(2023c)]%
        {Panoptic-wacv}
\bibfield{author}{\bibinfo{person}{Liyun Zhang}, \bibinfo{person}{Photchara Ratsamee}, \bibinfo{person}{Bowen Wang}, \bibinfo{person}{Zhaojie Luo}, \bibinfo{person}{Yuki Uranishi}, \bibinfo{person}{Manabu Higashida}, {and} \bibinfo{person}{Haruo Takemura}.} \bibinfo{year}{2023}\natexlab{c}.
\newblock \showarticletitle{Panoptic-aware image-to-image translation}. In \bibinfo{booktitle}{\emph{Proceedings of the IEEE/CVF winter conference on applications of computer vision}}. \bibinfo{pages}{259--268}.
\newblock


\bibitem[Zhang et~al\mbox{.}(2023d)]%
        {CNN-CM}
\bibfield{author}{\bibinfo{person}{Le Zhang}, \bibinfo{person}{Ryutaro Tanno}, \bibinfo{person}{Moucheng Xu}, \bibinfo{person}{Yawen Huang}, \bibinfo{person}{Kevin Bronik}, \bibinfo{person}{Chen Jin}, \bibinfo{person}{Joseph Jacob}, \bibinfo{person}{Yefeng Zheng}, \bibinfo{person}{Ling Shao}, \bibinfo{person}{Olga Ciccarelli}, {et~al\mbox{.}}} \bibinfo{year}{2023}\natexlab{d}.
\newblock \showarticletitle{Learning from multiple annotators for medical image segmentation}.
\newblock \bibinfo{journal}{\emph{Pattern Recognition}}  \bibinfo{volume}{138} (\bibinfo{year}{2023}), \bibinfo{pages}{109400}.
\newblock


\bibitem[Zhang et~al\mbox{.}(2023a)]%
        {MAGI}
\bibfield{author}{\bibinfo{person}{Yifei Zhang}, \bibinfo{person}{Siyi Gu}, \bibinfo{person}{Yuyang Gao}, \bibinfo{person}{Bo Pan}, \bibinfo{person}{Xiaofeng Yang}, {and} \bibinfo{person}{Liang Zhao}.} \bibinfo{year}{2023}\natexlab{a}.
\newblock \showarticletitle{Magi: Multi-annotated explanation-guided learning}. In \bibinfo{booktitle}{\emph{Proceedings of the IEEE/CVF International Conference on Computer Vision}}. \bibinfo{pages}{1977--1987}.
\newblock


\end{thebibliography}

\end{document}